\documentclass[aps,showpacs,nofootinbib]{revtex4-2}
\usepackage{graphicx}
\usepackage{bm}
\usepackage{graphicx}
\usepackage{amsmath}
\usepackage{mathrsfs}
\usepackage{ulem}
\usepackage{color}
\usepackage[colorlinks, linkcolor=red,anchorcolor=green,citecolor=blue]{hyperref}

\newtheorem{theorem}{Theorem}

\newcommand{\nn}{\nonumber}
\newcommand\diag{\operatorname{diag}}

\graphicspath{{./figs/}}

\begin{document}  
	
\title{Relaxation time approximation revisited and non-analytical structure in retarded correlators}

\author{Jin Hu}
\email{hu-j23@fzu.edu.cn}
\affiliation{Department of Physics, Fuzhou University, Fujian 350116, China}

\begin{abstract}
In this paper, we give a rigorous mathematical justification for the relaxation time approximation (RTA) model. We find that only the RTA with an energy-independent relaxation time can be justified in the case of hard interactions. Accordingly, we propose an alternative approach to restore the collision invariance lacking in traditional RTA. Besides, we provide a general statement on the non-analytical structures in the retarded correlators within the kinetic description. For hard interactions, hydrodynamic poles are the long-lived modes. Whereas for soft interactions, commonly encountered in relativistic kinetic theory, the gapless eigenvalue spectrum of linearized collision operator leads to gapless branch-cuts. We note that particle mass and inhomogeneous perturbations would complicate the above-mentioned non-analytical structures.

\end{abstract}




\maketitle	
\section{Introduction}\label{s1}
Over the past several decades, relativistic hydrodynamics has notably contributed to characterizing the dynamic evolution of Quark-Gluon Plasma (QGP), as observed in heavy-ion collision experiments at the Relativistic Heavy Ion Collider (RHIC) and the Large Hadron Collider (LHC) \cite{Romatschke:2017ejr}. Within the framework of phenomenological modeling for these collisions, the initialization time of hydrodynamic simulation is usually set to be less than $1$ fm/c  to match experimental observations. This  suggests that the system may rapidly equilibrate.  Moreover, the applicability of hydrodynamics  has been noted to extend into less expected domains, including small collision systems like nucleus-nucleon and proton-proton collisions \cite{CMS:2016fnw,ATLAS:2017hap}, which implies that these small collision systems could possibly exhibit fluid-like behaviors. What is the dynamic mechanism that triggers rapid equilibration? Why does relativistic hydrodynamics also work well for small systems? Regarding the first question, it has been proposed that hydrodynamics can be viewed as an attractor governing the late-time behavior of systems as they approach equilibrium \cite{Heller:2015dha}. At the early stage of evolution, the system tends to flow towards the hydrodynamic attractor, even when far from local equilibrium, which may account for rapid hydrodynamization. Thus, hydrodynamization may have a broader application range than thermalization, contrasting with the conventional view that hydrodynamics is a truncated gradient expansion near local equilibrium. Despite significant advancements, it remains an open question how hydrodynamization emerges from a general dynamic system with diverse microscopic interactions.

As the first step towards understanding how relativistic nonequilibrium systems reach thermal equilibrium, the properties of retarded correlation functions have recently garnered  extensive research interests. Two-point retarded correlation functions are pivotal, as they provide a wealth of insights into the transport characteristics of multi-particle systems, particularly how an equilibrium system responds to off-equilibrium disturbances within the linear regime. Moreover, non-analytical structures—such as poles or discontinuities in Fourier space—are crucial for determining the system's evolutionary patterns. Poles are indicative of collective excitations that evolve towards equilibrium, corresponding to the persistent hydrodynamic modes, whereas the presence of  non-hydrodynamic modes, which could be cuts or poles, is intrinsically linked to the emergence of hydrodynamic phenomena and the applicability of hydrodynamics. Research has shown  that the correlators contain only poles at infinite 't Hooft coupling in large $N$ thermal  $\mathcal{N} =4$ Super Yang-Mills (SYM) theory. In the holographic description, the spectra of correlation functions correspond to the ring-down spectra of dual linearly perturbed black holes: the quasinormal modes \cite{Hartnoll:2005ju,Kovtun:2005ev}.  However, in Ref. \cite{Hartnoll:2005ju}, the authors argue that the branch-cut structure emerges in the regime of weak, but finite 't Hooft coupling, indicating a possible transition behavior controlled by the 't Hooft coupling. 
 Motivated by this illuminating finding, Paul Romatschke calculated thermal correlators of large $N$ gauge theories in effective kinetic theory \cite{Romatschke:2015gic} (see also \cite{Hong:2010at,Bajec:2024jez}), reporting the onset of transition behavior for hydrodynamic poles. In other related studies, including analytical estimates \cite{strain2010}, qualitative models \cite{Kurkela:2017xis,Brants:2024wrx}, and numerical calculations \cite{Moore:2018mma,Ochsenfeld:2023wxz}, the dominant non-analytical structure is found to be the branch-cut rather than poles, thus posing the 'poles or cuts?' dilemma. Specifically, this dilemma initially refers to the mathematical essence of non-hydrodynamic excitations. In this work, we also use it to denote the comparison of  the lifetime of hydrodynamic/non-hydrodynamic excitations. 
 Also, it is worth noting that in a recent study \cite{Carballo:2024kbk} on the transient dynamics of quasinormal modes within a holographic framework, the authors demonstrated the existence of arbitrarily long-lived excitations arising from summation of short-lived quasinormal modes. These correspond to initial conditions in which a packet of energy is localized near the future horizon, a feature that bears a resemblance to the numerical results presented in \cite{Moore:2018mma,Ochsenfeld:2023wxz}. This may suggest a deeper and more direct connection between the linear response behaviors described by weakly-coupled kinetic theory and strongly-coupled holographic theories.
 

As a universal low-energy effective theory,  hydrodynamics describes  the collective, macroscopic dynamics over large distances and time scales. As previously mentioned, elucidating the relationship between the microscopic dynamics of matter's fundamental constituents and the macroscopic dynamics of coarse-grained degrees of freedom is a fundamental question. Typically, collective macroscopic dynamics heavily relies on the intricate microscopic dynamic details involving a vast number of degrees of freedom:  for instance, a microscopic theory is required to determine the relevant coefficients in the macroscopic description. In principle,  an accurate description of a dynamic system necessitates a comprehensive treatment of all particles at the microscopic level. However, tracking the evolution of all particles is impractical; thus, a reduced description of microscopic degrees of freedom is essential.  In scenarios of weak coupling, where the concept of quasiparticles is applicable, kinetic theory serves as an effective tool for describing many-body systems, with the relevant degrees of freedom being statistical distribution functions. For example, the BBGKY hierarchy equation describes the coupled dynamic evolution of n-particle distribution functions \cite{Bogoliubov,green,Kirkwood,Yvon} .  The BBGKY hierarchy equation is an infinite tower of integro-differential equations, making it challenging to solve without  truncation. By disregarding n-particle correlations (for $n \geq 2$) and truncating this infinite series of integro-differential equations to the lowest order,  the renowned Boltzmann equation is recovered, which describes the evolution of the one-particle distribution function. Despite its omission of nearly all particle correlations, the Boltzmann equation maintains profound physical significance and is capable of characterizing various non-equilibrium phenomena within kinetic theory. Notably, hydrodynamic behavior can be derived from the long-wavelength limit of the Boltzmann equation. Hence, the Boltzmann equation is an excellent candidate, particularly for weakly coupled cases, for elucidating how hydrodynamization is achieved in a dynamic system and how the multitude of microscopic degrees of freedom condense into coarse-grained hydrodynamic degrees of freedom.

However, the Boltzmann equation still contains an intractable collision integral, rendering its analytical or numerical solution challenging. Even with the simplest interactions, such as the hard-sphere potential, the linearized Boltzmann equation's collision operator retains a complex structure, making its analysis extremely difficult, 
see \cite{Wagner:2023joq} for related discussions on the linearized collision operator spectrum for hard-sphere interaction. Recently, the eigenspectrum of the linearized Boltzmann collision operator in massless scalar $\phi^4$ theory was analytically determined \cite{Denicol:2022bsq}. However, extending this to other realistic interactions remains a significant challenge\footnote{The first analytical eigenspectrum was obtained by C.S.Wang Chang and U.E.Uhlenbeck in the context of monatomic gases in the non-relativistic case (see chapter IV of \cite{deboer}).}. 
Marle \cite{Marlea,Marleb}, and Anderson and Witting (AW) \cite{1974Phy....74..466A} proposed an approximation of the relativistic Boltzmann equation using a simplified collision operator, effectively extending the BGK (Bhatnagar, Gross, and Krook) model to the relativistic domain \cite{bgk}. In the case of Marle, the proportionality factor is given by $m/\tau_R$, with $m$ the mass of the particles and $\tau_R$ the relaxation time, whereas for AW model, it is $u\cdot p/\tau_R$. We focus on the latter case in this work. The AW model, also known as the relaxation time approximation, omits much of the dynamic information  in the full collision operator. Firstly, it disregards nonlinearity. Secondly, its validity hinges on a clear separation between the eigenvalue representing the slowest relaxation and the others. Despite these limitations, the relaxation time approximation addresses practical challenges, enabling analytic and semi-analytic solutions for the simplified Boltzmann equation. This approach paves the way for an insightful and instructive analysis. 
Furthermore, the AW RTA model has been extended to a more general form with nontrivial energy dependence. However, the justification of this extended RTA model within kinetic theory awaits rigorous validation, which is another topic of interest. 

The  paper is organized as follows.  In Sec.~\ref{lke1}, we briefly review  the basic aspects of the  linearized Boltzmann equation. In Sec.~\ref{relax0},  we revisit the formulation of relaxation time approximation within the linearized Boltzmann equation. In this section, we provide a justification for the RTA, based on general mathematical considerations regarding the eigenspectrum structure of the linearized collision operator. It turns out that the RTA is well-justified exclusively in scenarios involving relativistic hard interactions. Sec.~\ref{relax1} serves as an application of the findings from Sec.~\ref{relax0} to address the non-analytical structures contained in the retarded correlation functions.  In Sec.~\ref{novel0}, we introduce a novel relaxation time approximation by truncating the full linearized operator, anticipating a broader range of applicability. 
Summary and outlook are given in Sec.~\ref{su}.  

Natural units $k_B=c=\hbar=1$ are employed. The metric tensor  is given by $g^{\mu\nu}=\diag(1,-1,-1,-1)$ , while $\Delta^{\mu\nu} \equiv g^{\mu\nu}-u^\mu u^\nu$ is the projection tensor orthogonal to the four-vector fluid velocity $u^\mu$. The abbreviation dP stands for $\int dP\equiv \frac{2}{(2\pi)^3}\int d^4p\, \theta(p^0)\delta(p^2 - m^2)$. 

~\\\textbf{Note}  --- 
Upon finalizing our manuscript, we became aware of a concurrent and highly pertinent study by L. Gavassino \cite{Gavassino:2024rck}, which also focuses on the  discussions about gapless modes, and has some overlap with our results.

\section{Linearized Boltzmann equation}
\label{lke1}
As the lowest order truncation of the relativistic BBGKY hierarchy, the on-shell relativistic Boltzmann equation describes the non-equilibrium evolution of a weakly coupled system,
\begin{align}
\label{Boltzmann}
&p\cdot\partial f(x,p)=C[f],\\
\label{Ckl}
C[f]
&\equiv\;
\int  {\rm dP^\prime}  {\rm dP_1} {\rm dP_2} \big(f(x,p_1)f(x,p_2)-f(x,p)f(x,p^\prime)\big) 
W_{p,p^\prime\to p_1,p_2},
\end{align}
where $f(x,p)$ is the one-particle distribution function in  phase space, and $C[f]$ represents the  collision kernel. Here  we neglect the external force and focus on local two-body collisions and the classical statistics. Furthermore, $W_{p,p^\prime\to p_1,p_2}=(2\pi)^6s\sigma(s,\Theta)\delta^{(4)}(p+p^\prime- p_1-p_2)$ with the differential differential cross section $\sigma(s,\Theta)$ encoding the interaction information. In subsequent discussions, the differential cross-section is alternatively expressed as $\sigma(g,\Theta)$ depending on $g\equiv \sqrt{-(p-p^\prime)\cdot (p-p^\prime)}$. Here $s$ represents the total center-of-momentum energy squared and $\Theta$ is the scattering angle in the  center-of-momentum frame.   The detailed balance property, $W_{p,p^\prime\to p_1,p_2}=W_{p_1,p_2\to p,p^\prime}$, inherent to the transition rates, is implicitly considered in the above expression.  Note that Eq.\eqref{Boltzmann} is specific to one-component systems; for generalizations to multi-component systems, refer to \cite{DeGroot:1980dk, Hu:2022vph}.


As mentioned before, the complicated Boltzmann equation is often linearized to  facilitate a more straightforward analysis.  Following the linearization procedure,  we expand the distribution function around the local equilibrium state $f(x,p)=f_{0}(x,p)(1+\chi(x,p)\,)$, then Eq.\eqref{Boltzmann} transforms into the following form,
\begin{align}
\label{boltz}
D  f(x,p)+E_p^{-1}p ^{\langle\nu\rangle}\partial_\nu f(x,p)=-f_{0}(x,p)\mathcal{L}_0[\chi],
\end{align}	 
with the linearized collision operator
\begin{align}
\label{cl}
-\mathcal{L}_0[\chi]&\equiv E_p^{-1}\int  {\rm dP^\prime}  {\rm dP_1} {\rm dP_2}f_{0}(x,p^\prime)W_{p,p^\prime\to p_1,p_2} \big(\chi(x,p_1)+\chi(x,p_2)-\chi(x,p)-\chi(x,p^\prime)\big),
\end{align}
where $D\equiv u\cdot\partial$, $E_p=u\cdot p$ and $p^{\langle\nu\rangle}\equiv \Delta^{\nu\rho}p_\rho$. 
The local equilibrium distribution is defined as
\begin{align}
\label{feq}
&f_{0}(x,p)=\exp[\xi(x)-\beta(x)\cdot p],
\end{align}
where $\beta^\mu\equiv\frac{u^\mu}{T},\xi\equiv \frac{\mu}{T},\beta\equiv\frac{1}{T}$ with the local temperature $T(x)$,  and  the chemical potential $\mu(x)$  associated with the conserved particle number. Evidently,  the collisional invariance of 1 and $p_{\mu}$ is respected by construction. From a mathematical standpoint, it can be verified that the linearized collision operator $\mathcal{L}_0$ is self-adjoint and positive semidefinite within the square-integrable Hilbert space \cite{DeGroot:1980dk},
\begin{align}
\label{inner0}
&\int  {\rm dP} f_{0}(p)E_p\psi(p)\mathcal{L}_0\phi(p)= \int  {\rm dP}f_{0}(p)E_p \phi(p)\mathcal{L}_0\psi(p),\nn\\
& \int  {\rm dP} f_{0}(p)E_p\psi(p)\mathcal{L}_0\psi(p) \geq 0,
\end{align}
where the spacetime dependence is neglected for simplicity.

In certain special cases, an additional simplification proves particularly useful.  When it comes to the  normal mode solution of the kinetic equation, the background around which the distribution function is expanded is typically assumed to be a homogeneous and static equilibrium configuration $f_{eq}(p)$. For the sake of subsequent discussions, we introduce this simplification as well. 
Adopting the expansion $f(x,p)=f_{eq}(p)(1+\chi(x,p)\,)$, Eq.(\ref{boltz}) can be further simplified to 
\begin{align}
\label{boltz2}
\partial_t \chi(x,p)+\bm{v}\cdot\nabla \chi(x,p)=-\mathcal{L}_0[\chi],
\end{align} 
where $\bm{v}\equiv \frac{\bm{p}}{p_0},\nabla_\alpha\equiv \Delta^\beta_\alpha\partial_\beta$, and we should  change $f_0(x,p)$ into $f_{eq}(p)$ in the above expressions accordingly.




\section{ Revisiting the  relaxation time approximation}
\label{relax0}
The relaxation time approximation of the  Boltzmann equation, known as the BGK model  in the nonrelativistic case and the AW model in the relativistic case, offers an effective description within kinetic theory. Notably, its simple mathematical structure facilitates the analytical extraction of underlying physics, albeit at the cost of precision. In this section, we examine the relationship between the RTA model and its complete form, the (linearized) Boltzmann equation, to elucidate the approximation's limitations, specifically when the model is valid within the framework of linearized kinetic theory. Note as an aside, the relaxation time is allowed to possess power-law energy dependence, conveniently parameterized as follows \cite{Dusling:2009df,Dusling:2011fd,Kurkela:2017xis}
\begin{equation}
\label{tauR}
\tau_R = (\beta E_p)^\alpha t_R.
\end{equation}
Here, $\alpha$ is an arbitrary constant controlling the energy dependence of the relaxation time, while $t_R$ is independent of momentum.  The specific value of $\alpha$ is believed to depend on the dynamic details and corresponds to various physical scenarios: $\alpha=0$ corresponds to the traditional AW RTA \cite{1974Phy....74..466A}; $\alpha=0.38$ is argued to well approximate the effective kinetic descriptions of quantum chromodynamics \cite{Rocha:2021zcw,Dusling:2009df,Dusling:2011fd}; while $\alpha=0.5$ is a good modeling in extreme out-of-equilibrium perturbations, e.g., jets, in this case $\tau_R$ is related to the famous jet stopping time \cite{Baier:1996kr,Baier:1996sk}. The successes achieved through flexible parameterization suggest that the RTA may effectively model realistic scenarios in practical applications.  Irrespective of phenomenological considerations from practical simulations, we concentrate solely on the mathematical aspects of the RTA model. We demonstrate below that the energy dependence is ascribed to the redefinition of the linearized collision operator
\begin{align}
\label{boltz11}
D  f(x,p)+E_p^{-1}p ^{\langle\nu\rangle}\partial_\nu f(x,p)=-f_{0}(x,p)E_p^{-\alpha}\mathcal{L}_\alpha[\chi],
\end{align}	
where $\mathcal{L}_\alpha\equiv E_p^\alpha \mathcal{L}_0$ represents the redefined  linearized collision operator with $\mathcal{L}_1$ and $\mathcal{L}_0$ being specific instances.  It is easily proved that $\mathcal{L}_\alpha$ inherits the positive semidefinite and self-adjoint properties from $\mathcal{L}_0$.  This can be achieved by redefining the weight function within the inner product definition as  $f_0(x,p)E_p^{1-\alpha}$ in Eq.(\ref{inner0}), which means the square-integrable function space should be also altered accordingly. 
It should  be noted that the corresponding integrals within the definitions should converge, which excludes very large values for $\alpha$.
In the following subsections  \ref{relax02} and \ref{relax03}, we illustrate that naively truncating $\mathcal{L}_\alpha$ leads to a form resembling the energy-dependent RTA; however the resulting RTA-like  model is not well-defined.

\subsection{ Anderson and Witting model}
\label{relax01}
The traditional RTA proposed by  Anderson and Witting is a relativistic generalization of the BGK model
\begin{align}
\label{aw}
p\cdot \partial f(x,p)=-\frac{E_p}{\tau_R}(f(x,p)-f_0(x,p)\,),
\end{align}	 
where $\tau_R$ is energy-independent,  corresponding to $\alpha = 0$ in Eq.\eqref{tauR}. The above equation can be rewritten as 
\begin{align}
\label{aw1}
D  f(x,p)+E_p^{-1}p ^{\langle\nu\rangle}\partial_\nu f(x,p)=-f_0(x,p)\frac{1}{\tau_R}\chi(x,p).
\end{align}
By substituting $\mathcal{L}_0[\chi]$ with $\frac{1}{\tau_R}\chi$ in Eq.\eqref{boltz}, effectively treating $\mathcal{L}_0$ as an identity operator (up to a constant factor), Eq.\eqref{boltz} simplifies to Eq.\eqref{aw1}. Note we have specified  the linearized collision operator $\mathcal{L}_0$ as our focus in this subsection. As will be manifest, $\frac{1}{\tau_R}$ can be identified as the smallest eigenvalue of $\mathcal{L}_0$.

To elucidate this, by linearizing the Boltzmann equation around a stationary homogeneous distribution $f_{eq}(p)$ and focusing on spatially uniform transport, we arrive at the equation
\begin{align}
\label{aw2}
\partial_t \chi(t,p)=-\mathcal{L}_0[\chi],
\end{align}	
which can be formally solved to get $\chi(t)=e^{-\mathcal{L}_0t}\chi(0)$. Since $\mathcal{L}_0$ is self-adjoint and semipositive, the linear perturbation from equilibrium decays exponentially, which is consistent with stability requirements. The same working logic applied to Eq.(\ref{aw1}) leads to $\chi^\prime(t)=e^{-t/\tau_R}\chi^\prime(0)$. By comparison, it is natural to interpret $\tau_R$ as the relaxation time for the linear perturbation from equilibrium, and $\mathcal{L}_0^{-1}$ as the infinite-dimensional relaxation time matrix (we adopt the terminology “matrix” although it may not be appropriate to use it when the eigenvalue spectrum is continuous).

Recognized as a self-adjoint and semipositive operator in Hilbert space,  $\mathcal{L}_0$ may contain both discrete and continuous eigenvalues. By expressing $\chi$ as a linear combination of the eigenfunctions ${\psi_n}$ of $\mathcal{L}_0$, 
\begin{align}
\chi(t,p)=\sum_{n} c_n(t)\psi_n(p),\quad \mathcal{L}_0 [\chi]=\sum_{n}\gamma_nc_n(t)\psi_n(p)
\end{align}
where $\sum_{n}$ implicitly includes the integral over the continuous eigenvalue spectrum, and we use $\gamma_n$ to denote  the eigenvalues of $\mathcal{L}_0$. With the eigenfunction expansion, Eq.(\ref{aw2}) can be resolved to reach
\begin{align}
\label{chi1}
\chi(t,p)&=\sum_{n} c_n(0)\psi_n(p)e^{-\gamma_n t}+C(p)=\sum_{n>5} c_n(0)\psi_n(p)e^{-\gamma_n t}+\sum_{n=1}^{5}c_n(0)\psi_n(p)+C(p),\\
\label{chi2}
\chi^\prime(t,p)&=\sum_{n} c^\prime_n(0)\psi_n(p)e^{-t/\tau_R}-\sum_{n=1}^5 c^\prime_n(0)\psi_n(p)e^{-t/\tau_R}+C^\prime(p)=\sum_{n>5} c^\prime_n(0)\psi_n(p)e^{-t/\tau_R}+C^\prime(p).
\end{align}
For comparative purposes, we also present the expansion form of $\chi^\prime = \sum_{n} c^\prime_n(t)\psi_n(p)$. Additionally, the counter term $-\sum_{n=1}^5 c^\prime_n(0)\psi_n(p)e^{-t/\tau_R}$ is introduced to account for the absence of collision invariance in the traditional RTA, as further discussed in Sec.\ref{novel0}. Note $C(p), C^\prime(p)$ are  integration constants. In the above equations, the eigenvalues are sequenced in ascending order when increasing $n$. As clearly seen in Eq.(\ref{chi1}), the collision invariants, $\psi_n (n=1,\cdots 5)$ with zero eigenvalues $\gamma_n=0$, contribute to the deviation function $\chi$ but remain unchanged with time. In the solution of Eq.\eqref{chi1}, the second term, reflecting the contribution from collision invariants, can also be regarded as an integration constant. By comparing Eqs.\eqref{chi1} and \eqref{chi2}, we can deduce that the traditional RTA simplifies the model by condensing the entire nonzero eigenvalue spectrum into a single representative eigenvalue, ${\gamma_n} \rightarrow \frac{1}{\tau_R}$. Given the exponentially decaying form, $1/\tau_R$ should be identified as the smallest eigenvalue, where the mode $\psi_6$ persists until the late stage of evolution. In other words, other modes with a larger eigenvalue, which should have been absent from an earlier time,  extend their lifetime to the final stage in this approximation.

However, the RTA model is justified only when the eigenvalue sequence ${\gamma_n}$ is distinctly separated from the origin. If this is not the case, such as when the eigenvalue spectrum extends continuously from the origin to other nonzero points, the relaxation time cannot be defined as the inverse of the smallest eigenvalue, as it might diverge,
leading to the elimination of collisional effects. 
Thus, the justification of the RTA heavily depends on the eigenspectrum structure of the linearized Boltzmann collision operator, particularly the characteristics of the eigenspectrum near the origin. As shown in \ref{relax04} of this section, the traditional RTA is permissible only in scenarios of 'hard-interaction' collisions, where the eigenspectrum of the linearized collision operator features a continuous range from $\nu_0\,(\nu_0 > 0)$ to $\infty$, along with possibly some discrete points. In such scenarios, a gap exists between the origin and other nonzero eigenvalues, providing a basis for constructing the RTA.

\subsection{ Energy-dependent relaxation time approximation ($\alpha >0$)}
\label{relax02}

As noted in the convenient parameterization Eq.(\ref{tauR}), the relaxation time, the only parameter reflecting the microscopic dynamic details within the RTA model, is allowed to exhibit energy dependence. It has long been believed that incorporating various energy dependence can reveal characteristic features of bottom-up thermalization and uncover hidden aspects of the full kinetic description \cite{Kurkela:2017xis}. Furthermore, the linearized kinetic equation within the energy-dependent relaxation time approximation has been successfully applied to jet physics \cite{Baier:1996kr,Baier:1996sk}, hydrodynamic simulations \cite{ Dusling:2009df,Dusling:2011fd}, and Weyl semimetals \cite{Amoretti:2023hpb}. Does this imply that the model or approximation is well-justified? One  caution should be taken: an approximated model may capture some aspects of the underlying physics relevant to specific issues but may fail to do so for others. Nevertheless, elucidating how an approximation is   derived from a complete theory is always highly significant. In this subsection, we show mathematically how this model can be similarly derived from the linearized Boltzmann equation, and how an inconsistency appears hindering the sound justification of the model.

Before delving into a more general discussion,  let's consider the specific case where $\alpha = 1$ in Eq.(\ref{tauR}) for concreteness. The model thus becomes 
\begin{align}
\label{ed}
p\cdot \partial f(x,p)=-\frac{1}{\beta t_R}(f(x,p)-f_0(x,p)\,).
\end{align}
This can be further rewritten as
\begin{align}
\label{ed1}
D  f(x,p)+E_p^{-1}p ^{\langle\nu\rangle}\partial_\nu f(x,p)=-f_0(x,p)\frac{1}{\beta E_pt_R}\chi(x,p)
\end{align}
which can be compared with Eq.(\ref{boltz11}) but now the replacement of $\mathcal{L}_1 \rightarrow \frac{1}{\beta t_R}$ should be invoked. This specific example aligns precisely with the energy-dependent RTA utilized in \cite{Kurkela:2017xis,Denicol:2022bsq}.

Without delving into repetitive details, we now proceed to a general discussion concerning $\mathcal{L}_\alpha$ with positive $\alpha$, termed as $\mathcal{L}_{\alpha>0}$
\begin{align}
\label{chi3}
\chi(t,p)&=\sum_{n} d_n(0)\Psi_n(p)e^{-\gamma^\prime_n t/E_p^\alpha }+D(p)=\sum_{n>5} d_n(0)\Psi_n(p)e^{-\gamma^\prime_n t/E_p^\alpha}+\sum_{n=1}^{5}d_n(0)\Psi_n(p)+D(p),\\
\label{chi4}
\chi^\prime(t,p)&=\sum_{n} d_n^\prime(0)\Psi_n(p)e^{-t/(\beta E_p^\alpha t_R)}-\sum_{n=1}^5 d_n^\prime(0)\Psi_n(p)e^{-t/(\beta E_p^\alpha t_R)}+D^\prime(p)\nn\\
&=\sum_{n>5} d_n^\prime(0)\Psi_n(p)e^{-t/(\beta E_p^\alpha t_R)}+D^\prime(p),
\end{align}
where $\{\gamma^\prime_n,\Psi_n\}$ represents the eigensystem of $\mathcal{L}_{\alpha>0}=E^\alpha_p\mathcal{L}_0$, and $D(p),D^\prime(p)$ are integration constants. Given that $E_p^\alpha\geq 0$, it still holds that the eigen modes with a larger eigenvalue fade away more quickly. Depending on whether the eigenspectrum continuously extends to the origin, we decide to retain the  smallest eigenvalue but zero or not. It seems that the story  remains no changes at all compared to what has been done in the previous subsection.

However, that is not the case, as the relaxation timescale for eigenfunctions is $ E_p^\alpha /\gamma^\prime_n$: the perturbation contributed by the hard modes takes longer time to relax towards equilibrium. With a fixed $E_p$, approximating the linearized collision kernel with a single eigenvalue is valid, as it suffices to identify the slowest mode. If varying $E_p$, we may always encounter the following situation $E_p^\alpha/\gamma^\prime_6 < E^\alpha_{p^\prime}/\gamma^\prime_7$ ($p\ll p^\prime$), indicating that slower modes are excluded in the energy-dependent RTA. Hence, it's insufficient to consider only $\gamma^\prime_6$; we must account for an infinite series of $\gamma^\prime_n$. This renders the RTA derivation process questionable from a physical standpoint. Unlike AW RTA,  the proposed relaxation time, expressed as $E_p^\alpha / \gamma^\prime_6$, lacks an upper bound. Indeed, as $E_p^\alpha$ grows large  enough, the hierarchy among the eigenvalues ${\gamma^\prime_n}$ becomes irrelevant, suggesting that $\mathcal{L}_{\alpha>0}$ can not support an energy-dependent RTA, regardless of the interactions involved: we don't even have to discuss the eigenvalue spectrum structure of $\mathcal{L}_{\alpha>0}$.


\subsection{ Energy-dependent relaxation time approximation ($\alpha < 0$)}
\label{relax03} 

The lack of a well-defined  relaxation time is due to the lack of a gap:  As illustrated in Section \ref{relax02}, a bounded energy-dependent relaxation time cannot be identified therein, but we can take a lesson from Eq.(\ref{tauR}). It can be proved that such a gap can exist if the following three conditions are met:
\begin{itemize}\label{condition}
	\item 1. The redefined linearized collision operator has nonzero eigenvalues separated from the origin. 
	\item 2. The particles in consideration are  massive.  
	\item 3. $\alpha$ takes the negative value in the parameterization Eq.(\ref{tauR}).           	       	
\end{itemize}
The necessity of the third condition can be relaxed if we deviate from the widely used parameterization in Eq.(\ref{tauR}). Additionally, the second condition may also be loosened by employing alternative parameterizations. In this analysis, we focus our attention on the parameterization specified in Eq.(\ref{tauR}). 

Solving the corresponding linearized Boltzmann equation 
\begin{align}
\label{aw3}
\partial_t \chi(t,p)=-E_p^{-\alpha}\mathcal{L}_{\alpha<0}[\chi],
\end{align}	    
yields the formal solution effortlessly 
\begin{align}
\chi(t)=e^{-E_p^{-\alpha}t\mathcal{L}_{\alpha<0} }\chi(0).
\end{align}
We  “naively”  truncate $\mathcal{L}_{\alpha<0}$ to the smallest nonzero eigenvalue $\gamma^\prime_6$ (or the nonzero infimum if the spectrum is continuous) to reproduce Eq.(\ref{tauR}). Thus, the relaxation timescale for the perturbation decay is precisely $\frac{E_p^\alpha}{\gamma^\prime_6}$ bounded from above by $\frac{m^\alpha}{\gamma^\prime_6}$. This indicates that all modes attenuate at a finite rate with the slowest rate interpreted as the inverse relaxation time, $1/\tau_R=\frac{\gamma^\prime_6}{m^\alpha}$,  in contrast to what we encounter in  \ref{relax02}.  But if the particles are massless, then it follows that $\frac{E_p^\alpha}{\gamma^\prime_6}$ is  unbounded, and the well-defined RTA is also lacking for the similar reason shown in \ref{relax02}. 

However, we confirm that $\tau_R=\frac{m^\alpha}{\gamma^\prime_6}$  equates to the scenario where $\alpha=0$, i.e., the AW RTA, and does not introduce a new class of RTAs. This verification is ruled by the following physical argument. Suppose an observer perturbs the system in equilibrium with a small-amplitude spatially homogeneous disturbance at $t=0$, then the observer records when the disturbance dissipates to none. It is reasonable to require that Eqs.(\ref{boltz}), (\ref{boltz11}), and their corresponding RTAs (if any) give the same relaxation timescale.

Assuming that the eigenvalue spectrum of $\mathcal{L}_0$ is gapped,  the relaxation scale is given by $\frac{1}{\gamma_6}$ according to Eq.(\ref{boltz}). If we try to describe  the system using Eq.(\ref{boltz11}) with $\alpha < 0$, the relaxation scale is then dictated by the mode with the longest lifetime,  and reads $\tau_R(p\rightarrow 0)=\frac{m^\alpha}{\gamma^\prime_6}$. Hence at $t=\frac{m^\alpha}{\gamma^\prime_6}$, the observer announces that the system returns to its equilibrium state. Given that the same experiment is conducted for the same system, the observer should give identical observation results, as Eq.(\ref{boltz11}) is an equivalent transformation of Eq.(\ref{boltz}), i.e., $\frac{1}{\gamma_6} \simeq \frac{m^\alpha}{\gamma^\prime_6}$. For the $\alpha < 0$ case, even if there is an upper bound for relaxation scales and one can manufacture an energy-dependent (or precisely mass-dependent) RTA,  it is still a redundant description compared to AW RTA.   In summary, an energy-dependent RTA should not be regarded as a well-defined truncation to its UV complete theory, i.e., the linearized Boltzmann equation, irrespective of the interaction details. It is suggestive to keep only the description provided by $\mathcal{L}_0$, because it is simplest without energy-dependent factor, and the mathematical analysis specifically for  $\mathcal{L}_0$ is readily available \cite{dud,dud1}. 

The observer argument above 
can provide an equivalent justification for reconciling   the seemingly contradictory aspects between two detailed calculations for massless scalar $\phi^4$ theory with leading order interaction \cite{Ochsenfeld:2023wxz,Denicol:2022bsq}. Denicol and Noronha analytically calculate the eigenspectrum of $\mathcal{L}_1$ and obtain a series of discrete eigenvalues \cite{Denicol:2022bsq}. 
By utilizing the observer argument again, we can conclude that the eigenspectrum of $\mathcal{L}_0$ must extend continuously towards the origin, and thus gapless. This conclusion is consistent with the mathematical aspect: one can show that for leading order  scalar $\phi^4$ theory, $\sigma(g,\Theta)\sim \frac{1}{s}=\frac{1}{g^2+4m^2} < \frac{2}{g^2}$, so the interaction  falls within the soft interaction category according to the criteria given in \cite{dud,dud1}. Furthermore, the gapless continuous eigenspectrum has also been confirmed in recent calculations \cite{Ochsenfeld:2023wxz,Gavassino:2024rck,Rocha:2024cge}.

\subsection{ Hard interaction VS soft interaction}
\label{relax04}

Based on the above discussions, the Anderson and Witting model seems to be  the only candidate with a potentially well-defined justification. However, it's important to note that this justification is contingent upon the eigenspectrum properties near the origin, which are, in turn, dependent on the specifics of the interactions. In this subsection, we will demonstrate that the aforementioned interaction details pertain to the softness of the interaction: the interaction is classified as soft or hard based on the differential cross-section's form, as encapsulated in the following two mathematical theorems \cite{dud,dud1}. Before presenting these, let's briefly review the basic aspects.  
Following \cite{dud,dud1}, we linearize the Boltzmann equation around the global equilibrium distribution. But  quite differently, the distribution function is written as 
\begin{align}
\label{linear}
f(x,p)=f_{eq}(p)+f^{1/2}_{eq}(p)\tilde{\chi}(x,p),
\end{align} 
and the linearized Boltzmann equation is cast into
\begin{align}
\label{boltz02}
\partial_t \tilde{\chi}(x,p)+\bm{v}\cdot\nabla \tilde{\chi}(x,p)=-\mathcal{L}_0[\tilde{\chi}]\equiv-v(p)\tilde{\chi}(x,p)+K[\tilde{\chi}],
\end{align}	
with  
\begin{align}
\nu(p)&\equiv f^{1/2}_{eq}(p)p_0^{-1}\int  {\rm dP^\prime}  {\rm dP_1} {\rm dP_2}f_{eq}(p^\prime)W_{p,p^\prime\to p_1,p_2}\frac{ 1}{ f^{1/2}_{eq}(p)},\\
K[\tilde{\chi}] &\equiv  f^{1/2}_{eq}(p)p_0^{-1}\int  {\rm dP^\prime}  {\rm dP_1} {\rm dP_2}f_{eq}(p^\prime)W_{p,p^\prime\to p_1,p_2} \big(\frac{ \tilde{\chi}(x,p_1)}{ f^{1/2}_{eq}(p_1)}+\frac{ \tilde{\chi}(x,p_2)}{ f^{1/2}_{eq}(p_2)}-\frac{ \tilde{\chi}(x,p^\prime)}{ f^{1/2}_{eq}(p^\prime)}\big),
\end{align}
where $\nu(p)$ represents the collision frequency and we introduce a compact operator $K$.
Here we maintain the same notations as in the previous context to avoid complicating the overall notation system, although the expansion Eq.(\ref{linear}) looks quite different. Furthermore, it is immediately observable that  Eq.(\ref{boltz02}) shares the similar form as Eq.(\ref{boltz2}). As a result, they should also share the similar formal solution $\tilde{\chi}(t)=e^{-\mathcal{L}_0t}\tilde{\chi}(0)$ when the system is homogeneously perturbed.  In both cases, whether considering  Eq.(\ref{boltz02}) or Eq.(\ref{boltz2}), $\chi$ or $\tilde{\chi}$ should be regarded as the perturbation above the same equilibrium state. Given that both Eqs.\eqref{boltz02} and \eqref{boltz2} are derived from the same Boltzmann equation through linearization, we would not expect the eigenspectrum for $\mathcal{L}_0$ in Eq.\eqref{boltz02} to differ from that in Eq.\eqref{boltz2}. If this were the case, the decay behavior of the perturbation would exhibit significant differences, signifying the physical inconsistency. This is another successful application of the observer argument from the previous subsection. Therefore, we argue that the eigenspectrum of $\mathcal{L}_0$ should be identical  in Eqs.(\ref{boltz02}) and (\ref{boltz2}), and this is why we maintain the same notations. Then let's come to the core of this subsection:

\begin{theorem}
	\begin{itemize}
		\item Assume that $\exists \gamma > -2, \, 0\leq \beta < \gamma+2,\, B>0$ and $c_0 >0 $ , so that $\sigma(g,\Theta) > B\frac{g^{\beta+1}}{c_0+g}\sin^\gamma\Theta$, then $\nu(p) > \nu_0 (p_0/m)^{\beta/2}$ where $\nu_0$ is a constant, the interaction is hard.
	\end{itemize}
\end{theorem}
\begin{theorem}
	\begin{itemize}
		\item Assume that $\exists 0 < \alpha < 4,\, \gamma> -2$ and $B^\prime > 0$, so that $\sigma(g,\Theta) < B^\prime g^{-\alpha}\sin^\gamma\Theta$, then $\nu(p) < \nu_0 (p_0/m)^{-\epsilon/2} \leq \nu_0$, the interaction is soft, where
		\begin{align}
		\epsilon=
		\begin{cases}
		\displaystyle
		\alpha, \quad \text{for} \quad 0 < \alpha < 3,
		\\
		\alpha-2, \quad \text{for} \quad 3 < \alpha < 4,\\
		\delta +1, \quad \text{for} \quad \alpha = 3, \quad \text{and} \quad  0 < \delta < 1,
		\end{cases}
		\end{align}
		and $\nu_0$ is a constant,
	\end{itemize}
\end{theorem}
where the proofs of these theorems can be found in \cite{dud1}.

Analyzing the collision frequency $\nu(p)$ enables us to distinguish between relativistic soft and hard interactions.  Mathematical analysis indicates that relativistic interactions tend to be softer than their nonrelativistic counterparts \cite{dud,dud1}. It can be also proved that for soft interactions $-\mathcal{L}_0$ is a bounded operator with the eigenspectrum $[-\nu_{max},0]$, whereas for hard interactions, $-\mathcal{L}_0$ is unbounded, featuring an eigenspectrum of $[-\infty,-\nu_{min}]$, where $\nu_{max}$ and $\nu_{min}$ can be estimated by the extrema of $\nu(p)$.
Given the eigenspectrum properties presented here, we conclude that the AW RTA is only well-justified in the case of hard interactions. For example, in the Weinberg-Salam theory at low energies \cite{DeGroot:1980dk}, e.g.,  four fermions interaction in the electroweak sector well below the gauge boson masses, the typical differential cross section behaves as $\sigma(g,\Theta)\sim s=g^2+4m^2 > \frac{g^3}{c_0+g} \geq \frac{g^3}{c_0+g}\sin\Theta$  with the conditions $\gamma=1,\beta=2, B=1$ in  Theorem 1. Consequently, the model of a gas consisting of elastically colliding neutrinos in the low-energy limit permits an RTA description of transport phenomena.  

As most relativistic interactions are soft according to Theorem 2, the AW RTA lacks a solid foundation in the majority of cases. For instance, in the  extensively studied scalar $\phi^4$ theory, the interaction is soft as exhibited in the previous section. This softness results in a branch-cut structure that extends across the entire negative imaginary axis in the retarded correlators \cite{Moore:2018mma,Perna:2021osw}. 

There is one comment left. When interactions are too complicated so that their classification as hard or soft cannot be easily known, we recommend employing the finite-element method described in \cite{Ochsenfeld:2023wxz} for analyzing  the eigenspectrum properties. By implementing the outlined procedures therein, we can establish the dictionary between various field theories and  the eigenspectrum structure within their linearized kinetic description. In this sense, we can exhaust  all commonly used  interactions admitting an RTA description.

\section{ Pole or cut\, --- the non-analytical structures in retarded correlators}
\label{relax1}

Two-point retarded correlation functions are crucial and insightful as they encapsulate rich information about the transport properties of many-body systems.   Their analytical structures can reflect the characteristic properties of how thermal equilibrium is reached. For instance,  the non-analytical structures --- poles or cuts in  Fourier space,  govern the evolution behavior of the system:  poles describe collective excitations evolving towards  equilibrium corresponding to  hydrodynamic modes, while the presence of cuts or non-hydrodynamic modes is closely linked to the emergence of hydrodynamic behavior and the applicability of hydrodynamics. Thus, the research into the analytical properties of retarded correlators is profound, which is initially explored by Romatschke in the weakly-coupled kinetic  theory  \cite{Romatschke:2015gic}. Romatschke's findings highlight two key features in the analytical properties of the retarded correlators: the cuts are gapped corresponding to non-hydrodynamic modes, below which hydrodynamic poles dominate as long-lived degrees of freedom;  the hydrodynamic poles cease to exist for some critical value of the wavenumber reminiscent of the phenomenon of onset transitions, which are successfully reproduced within the mutilated RTA model detailed in \cite{Hu:2023elg}. As a supplement, we note that the universal behavior of onset transitions has been reported for a long time in the context of the nonrelativistic kinetic theory using the  mutilated model \cite{deboer}. However, the AW RTA adopted in \cite{Romatschke:2015gic} is less well-founded than its nonrelativistic counterpart, given that interactions tend to be softer in relativistic scenarios: the universal onset transitions in nonrelativistic systems may be rarely observed in relativistic systems due to varying degrees of interaction softness.

Later on, Kurkela and Wiedemann reexamined the behavior of the retarded correlators, beginning with the parametrized energy-dependent RTA as described in Eq. \eqref{tauR}, with a particular focus on the case when $\alpha = 1$ \cite{Kurkela:2017xis}. Their conclusions, however, stand in contrast to those in  \cite{Romatschke:2015gic}:  there is no sharp onset of hydrodynamic behavior; the structure of cuts turns into the entire strip $\text{Im} \,\omega < 0, -k \leq \text{Re}\,\omega \leq k$ from a gapped line given in \cite{Romatschke:2015gic}. They also state that the appearance of poles in the first (physical) Riemann sheet of retarded correlation functions is a matter of choosing a particular analytical continuation and thus cannot be related unambiguously to the onset of fluid dynamic behavior. However, as demonstrated in  \ref{relax02}, their model corresponding to Eq.(\ref{tauR}) with $\alpha=1$ exhibits inconsistencies: it can only be seen as an incomplete truncation to its ultraviolet (UV) completion --- the linearized Boltzmann equation, because an infinite number of slow modes are excluded. Although the analysis given in \cite{Kurkela:2017xis} is still illuminating, we choose to work with RTA only in the hard interaction case for theoretical consistency.

As observed, these two studies introduced above correspond to the models discussed in \ref{relax01} and \ref{relax02}. In this place, we want to give a general statement on the topic of non-analytical structures in the (stress-stress) retarded correlators.  Before proceeding, let's elaborate on how to derive the non-analytical structure if the interactions are soft.  In this case,  the eigenvalue spectrum of $\mathcal{L}_0$ is gapless and the typical stress-stress retarded correlation function is given by
\begin{align}
G_R(\omega)=\int^{\nu_{max}}_0d\gamma \frac{\rho(\gamma)}{\omega+i\gamma}
\end{align}   
where we use $\gamma$ to denote the continuously distributed eigenvalues, and the weight function $\rho(\gamma)$ is nonzero in the integration range.
This expression can be derived from the Fourier transform of the linearized Boltzmann equation Eq.(\ref{boltz}) in the limit of vanishing $k$ \cite{Moore:2018mma,Perna:2021osw}.  This expression clearly exhibits discontinuity as we shift from $\omega = -i\gamma + \epsilon$ to $\omega = -i\gamma - \epsilon$, indicating the presence of a branch-cut line extending from $-i\nu_{max}$ to $0$. 

In the limit of vanishing mass and nonzero $k$, extracting the non-analytical structure becomes unmanageable.  
We first assume that $\cos\theta$, regarded as an operator in momentum-space functions, commute with $\mathcal{L}_0$,  then we can replace  $\omega$ by $\omega-k\cos\theta$, where $\theta$ is the angle between spatial components of $p$ and $k$. Then we can cast $G_R(\omega)$ into
\begin{align}
G_R(\omega,k)\sim \int_{-1}^1 d\cos\theta\int^{\nu_{max}}_0d\gamma \frac{\rho(\gamma)}{\omega-k\cos\theta+i\gamma},
\end{align}   
where the integration over $\cos\theta$ must be performed in the momentum integral. The resulting expression is
\begin{align}
G_R(\omega,k)\sim \frac{1}{k}\int^{\nu_{max}}_0d\gamma \rho(\gamma)\big(\log(\omega+k+i\gamma)-\log(\omega-k+i\gamma)\,\big).
\end{align}
This results in the branch-cut structure
\begin{align}
\label{nonzerok}
\text{Im }\omega=-\gamma, \quad -k \leq \text{Re}\,\omega \leq k,\quad 0 < \gamma < \nu_{max},
\end{align}
where  $\nu_{max}=\infty$ for a massless theory (see Eq.(\ref{blowup})\,). Ultimately, 
the resulting  branch-cut structure seemingly reproduces the result for nonzero $k$ given in \cite{Kurkela:2017xis} \footnote{ \cite{Kurkela:2017xis}  is based on the choice of $\mathcal{L}_1$, or more precisely, an incomplete truncation of $\mathcal{L}_1$, as shown in \ref{relax02}.  For completeness and consistency, a thorough study extending  \cite{Rocha:2024cge} to spatially inhomogeneous perturbations is warranted.}. 



However, a reminder should be given that Eq.(\ref{nonzerok}) relies heavily on the commutation approximation for $\cos\theta$ and $\mathcal{L}_0$, which breaks down when $\mathcal{L}_0$ is the complete linearized Boltzmann collision operator. In other words, the structure in Eq.(\ref{nonzerok}) should be modified. For situations involving hard interactions, the extent of modification should be minimal as the RTA can be seen as a good approximation to the complete linearized collision operator in that case.

Based on the above statement, determining the non-analytical structures in the retarded correlation functions equates to solving for the eigenspectrum of $\mathcal{L}_0$.  However, a contradictory question arises: both \cite{Moore:2018mma} and \cite{Ochsenfeld:2023wxz} demonstrate that the cut extends across the entire negative imaginary frequency axis, contrasting with the bounded region $[-i\,\nu_{max}, 0]$ inferred from mathematical analysis. The discrepancy vanishes because, in the context of a massless  theory with soft interactions, the collision frequency becomes unbounded  \cite{dud1}
\begin{align}
\label{blowup}
\nu(p)  \longrightarrow \infty,\quad \text{when}\quad p\rightarrow 0.
\end{align}
Therefore, $\nu_{max}=\infty$, and these results are consistent with each other.

For clarity, we give a summary of this section by reconsidering the interplay of  nonzero particle mass $m$ and wavenumber $k$,
and our statement can be summarized as follows:
\begin{itemize}
	\item Hard interactions: the RTA is a well-defined approximation relative to its UV complete theory.
	\begin{itemize}
		\item $m=0$ or $k=0$: Romatschke's analysis applies: the retarded correlators (including stress-stress correlator) exhibit a gapped branch-cut line (only two endpoints are branch points)
		\begin{align}
		\label{branch}
		\text{Im }\omega=-\frac{1}{\tau_R}, \quad -k \leq \text{Re}\,\omega \leq k,
		\end{align}
		associated with nonhydrodynamic modes. Below this gap,  there is a window where hydrodynamic modes become the long-lived degrees of freedom.
		\item $m\neq 0,\, k\neq 0$: The non-analytical structures of the retarded correlators become complicated due to the interplay between the nonzero particle mass $m$ and wavenumber $k$.  Upon inspecting the derivation detailed in \cite{Romatschke:2015gic}, we find that if the particles are massive, the free-streaming term is  proportional to $\bm{v}\cdot \nabla \sim \frac{pk\cos\theta}{\sqrt{p^2+m^2}}$ depending on $p$. This could introduce additional complex structures into correlators after performing momentum integral, beyond the poles or gapped cuts predicted by Romatschke. We leave this discussion to  future work. For now, we can only conclude that the non-analytical structure remains in the form of Eq.(\ref{branch}). The remarkable distinction is that the points lying within the branch-cut line are all branch points.  
	\end{itemize}        	
	\item Soft interactions. The RTA does not constitute a well-defined model. The simplifying assumptions should at least retain the eigenvalues and eigenfunctions of $\mathcal{L}_0$ near the origin, specifically within the region $[0,\nu_{max}]$. 
	\begin{itemize}
		\item $k=0$: In this case, the stress-stress correlator possesses a branch cut described by
		\begin{align}
		\label{gye}
		-\nu_{max}<\text{Im }\omega < 0, \quad  \text{Re}\,\omega =0,
		\end{align}
		which in the limit of vanishing mass matches the conclusion drawn in \cite{Moore:2018mma}. Strictly speaking, the author focuses on the discussions on the symmetrized 2-point function therein, but the retarded correlation function can be related to it through KMS relation. At weak coupling where the relevant frequencies will be suppressed by powers of the coupling, the result aligns with Eq.(\ref{gye}), which is further confirmed
		in a related study discussing $\mathcal{L}_1$ \cite{Rocha:2024cge}.	        	
		\item $k\neq 0$: The complication arising from $ \frac{pk\cos\theta}{\sqrt{p^2+m^2}}$ also exists in this scenario, and the issue should be trickier than the case of massive RTA. Even if $m=0$,  it is still difficult to reach a concise-form conclusion due to the non-commutativity  of $\cos\theta$ and $\mathcal{L}_0$. Therefore, the non-analytical structures don’t exhibit regular shapes like Eq.({\ref{nonzerok}}).
	\end{itemize}
	
\end{itemize}

In the end of this section, we would like to add several comments  as follows:

--- Although failing to reach a definite conclusion in the cases involved with nonzero $k$ or/and $m$, we plan to numerically solve the issue following finite-element analysis given in \cite{Ochsenfeld:2023wxz}, which is left to future work. For instance, for soft interactions, we can examine the simplest interaction case, scalar $\phi^4$ theory with leading-order interaction, which is expected to reveal the property of non-analytical structures qualitatively. 

--- If Eq.(\ref{nonzerok})  holds, hydrodynamic modes are completely embedded in this strip, leading to the conclusion that nonhydrodynamic modes related to free-streaming dynamics are long-lived degrees of freedom compared to hydrodynamic modes. According to our proposed non-commutativity between $\cos\theta$ and $\mathcal{L}_0$, Eq.(\ref{nonzerok})  would be modified so that there could be a window where hydrodynamic modes dominate as long-lived modes. In this sense, the physical picture may change by reconsidering the impact of the non-commutativity. 

--- In this script (likewise in \cite{Gavassino:2024rck}), we focus on the gapped/gapless property of nonhydrodynamic modes.  By using the terminology “dominant" or “dominate", we refer to the comparison of their lifetime. In practical cases, the dominant role should be assessed by combining their corresponding residues or discontinuity, which encode their contribution to the retarded correlation functions. This issue can be addressed using numerical calculations outlined in the first comment.


\section{The novel relaxation time approximation}
\label{novel0}

This section focuses on the case of hard interactions, where the relaxation time approximation is validated.  Despite its validation, the relaxation time approximation by Anderson and Witting still has inherent flaws as it fails to respect the collision invariance of the Boltzmann collision operator, which is tantamount to microscopic conservation laws.  To address this issue, the counter terms to restore collision invariance are introduced,
\begin{align}
\label{L}
-\mathcal{L}_0\simeq \big(-\gamma_6+\gamma_6\sum_{n=1}^{5}|\psi_n\rangle\langle \psi_n|\big),
\end{align} 
which is referred to as the novel RTA in \cite{Denicol:2022bsq,Rocha:2021zcw}, and is also known as the mutilated  operator in \cite{Hu:2023elg,Hu:2022xjn,Hu:2022mvl} (see chapter V of \cite{deboer} for an earlier discussion). Here $|\psi_n\rangle$ denotes the orthonormal eigenfunctions of $\mathcal{L}_0$ with zero eigenvalues, and $\gamma_6$ represents the smallest positive eigenvalue with the dimension of $[E]$. Thus we recover the collision invariance: $\mathcal{L}_0|\psi_n\rangle=0, n=1,\cdots 5$. To relate it to the traditional  RTA, one needs to  identify the relaxation time as $\tau_R\equiv\frac{1 }{\gamma_6}$. 
When $\mathcal{L}_0$ acts on other eigenfunctions, it results in $\mathcal{L}_0|\psi_n\rangle = \gamma_6 |\psi_n\rangle$ for $n > 5$, collapsing all positive eigenvalues into the smallest positive one. This is why the model in Eq.\eqref{L} is sometimes referred to as "mutilated" \cite{deboer}.

Macroscopic conservation laws are inherently restored when microscopic conservation laws, specifically the collision invariance of the collision operator, are respected in the model's construction, thereby simultaneously fixing the basic flaws. The novel RTA provides the flexibility to adjust the matching conditions used in kinetic theory, a feature not present in the traditional RTA as discussed in \cite{Peralta-Ramos:2012tgz,Rocha:2021zcw,Rocha:2022fqz,Hu:2023elg,Kandus:2024tcv}.  This flexibility is particularly advantageous when discussing hydrodynamic frame dependence, such as in the first-order causal theory of the BDNK type \cite{Bemfica:2017wps,Kovtun:2019hdm,Bemfica:2019knx,Bemfica:2020zjp}.  


By specifying the inner product definition, we can cast Eq.\eqref{L} into a less abstract form. For brevity, we omit the $x$ dependence in this section. Omitting the derivation details (for which we refer readers to \cite{Rocha:2021zcw}), the novel RTA can be formulated as follows
\begin{equation}
\label{novel}
-\mathcal{L}_0\chi(p)=-\frac{1}{\tau _{R}}%
\left[ \chi(p)-\frac{ \left( 1, \chi(p)\right)}{\left( 1, 1\right)}-P^{(0)}_{1}\frac{\left( 1, P_{1}^{\left( 0\right) }\chi(p)\right)}{%
	\left( 1, P_{1}^{\left( 0\right) }P_{1}^{\left( 0\right) }\right)}-p^{\left\langle \mu \right\rangle }\frac{ \left( 1, p_{\left\langle
		\mu \right\rangle }\chi(p)\right)}{(1/3)
	\left( 1, p_{\left\langle \nu \right\rangle }p^{\left\langle \nu
		\right\rangle   }\right)}\right].
\end{equation}%
Here the orthogonal basis is
constructed from the collision invariants $1$ and $p^\mu$, and given by
\begin{equation}
P_{0}^{\left( 0\right) }\equiv 1,\left. {}\right. P_{1}^{\left( 0\right)
}\equiv 1-\frac{\left( 1,1 \right)}{\left( 1,E_p \right)} E_{p},\left. {}\right.
p^{\left\langle \mu \right\rangle }\equiv \Delta ^{\mu \nu }p_{\nu }\text{,}
\end{equation}%
and the definition for the inner product bracket is invoked
\begin{align}
\label{inner}
(B,C)\equiv\int  {\rm dP} w(p)B(p)C(p),
\end{align}
with the weight function 
\begin{align}
\label{weight0}
w(p)\equiv f_0(p)E_p,
\end{align}
note that the weight function is the same as the one  in Eq.(\ref{inner0}), as it should be.
In Eq.\eqref{novel}, the novel RTA is constructed by adding counter terms to the RTA, but it can also be derived through an alternative method, as shall be given below. To proceed, we expand $\chi(p)$ as the linear combination of the eigenfunctions of $\mathcal{L}_0$ as
\begin{align}
\chi(p)=\sum_{n}a_n\psi_n, \quad \text{with}\quad a_n\equiv \frac{(\chi(p),\psi_n)}{(\psi_n,\psi_n)},
\end{align}
where the summation can also denote the integral for continuous spectra. Then the action of $\mathcal{L}_0$ on $\chi$ leads to
\begin{align}
\label{noval1}
\mathcal{L}_0\chi (p)&=\sum_{n=1}^\infty a_n\gamma_n\psi_n\simeq \sum_{n=1}^{N}a_n\gamma_n\psi_n+\gamma_N\sum_{n=N+1}^\infty a_n\psi_n\nn\\
&=\sum_{n=1}^N a_n(\gamma_n-\gamma_N)\psi_n+\gamma_N\chi(p),
\end{align}
where as prescribed previously $\gamma_n$ is sequenced in ascending order, and we make the approximation in the first line \cite{Sirovich:1962}. If $N=6$, then the second term in the last line matches the RTA collision kernel. Given that $\gamma_i=0$ for $i$ ranging from $1$ to $5$, the above equation can be cast into 
\begin{align}
\mathcal{L}_0\chi (p)\simeq \gamma_6\chi(p)-\gamma_6\sum_{n=1}^5 a_n\psi_n
\end{align}
where the second term on the right-hand side precisely constitutes the counter term required to restore collision invariance. Hence, Eq.\eqref{noval1} encompasses Eq.\eqref{novel} as a particular case. One can also show that the collision invariance is restored for $N>6$. Compared to the constructions in \cite{Rocha:2021zcw,Hu:2023elg,Hu:2022mvl}, the method introduced here embodies the essence of truncation that leads to RTA and is capable of accommodating a broader spectrum of modified RTAs. By increasing $N$, we can incorporate additional eigenvalues and eigenfunctions into the original RTA if available. This approach is beneficial when our knowledge is limited to a finite set of eigenvalues and eigenfunctions (these can be determined numerically at  least \cite{Ochsenfeld:2023wxz}), when we attempt to integrate the relevant information into the RTA. Furthermore, the model in Eq.\eqref{noval1}, serving as an intermediate between the RTA and the full linearized collision operator, can be fine-tuned to balance accuracy and simplicity, allowing for a judicious compromise.


\section{Summary and outlook}
\label{su}
In this paper, we revisit the widely used relaxation time approximation within the linearized Boltzmann equation. According to the mathematical analysis on the eigenspectrum of linearized Boltzmann collision operator $\mathcal{L}_0$, the RTA model is justified only for hard interactions, thereby  ruling out the energy-dependent parametrization in Eq.\eqref{tauR}.
The consideration is grounded on mathematical aspects, and Eq.(\ref{tauR}) can effectively serve as a convenient parameterized model. Furthermore, Eq.(\ref{tauR}) sheds light on the redefinition of  linearized collision operator denoted as $\mathcal{L}_\alpha$. 
We also provide a derivation of the novel RTA to restore the collision invariance by truncating the linearized collision operator, instead of by adding counter terms.

For discussing the non-analytical structures within the retarded correlators, we find that focusing on $\mathcal{L}_0$ and its eigenspectrum properties is the simplest approach. When interactions are hard, the RTA is well-defined, and the analytical properties of retarded correlators, as detailed in \cite{Romatschke:2015gic}, are applicable. There is a gap between the branch-cut lines  and  the real axis $\text{Im}\, \omega=0$. Therefore, the gapless hydrodynamic modes are well-defined low-energy degrees of freedom when $k$ is small. However, according to mathematical derivations, relativistic interactions are often soft, leading us to focus on the alternative scenario in most cases. In scenarios with soft interactions, the RTA is no longer well-justified, and the dominating long-lived non-analytical structure turns into the branch-cut or the non-hydrodynamic modes. Our conclusion is consistent with the previous related studies with the comparison details elaborated in the main text. Note that if the particles constituting the system are massive or the perturbations are inhomogeneous, the non-analytical structures would be more complicated and richer. 

There are possible extensions to the present research. As mentioned in \ref{relax04}, we can establish a dictionary between various field theories and the eigenspectrum structure within their linearized kinetic description. This can help us determine the dominant non-analytical structure and derive all possible RTA models. The former concerns the properties of the retarded correlation function of many-body systems, while the latter provides a solid theoretical basis for RTA, if any. We believe that applying RTA to a justified system is theoretically more consistent than doing so without justification, which is one of the motivations for this work. Additionally, almost all related research on the properties of retarded correlators is based on the linearized kinetic description. It would be interesting to explore the impacts of the nonlinear structure contained in the complete kinetic description, such as the Boltzmann equation, on the present conclusions.

\section*{Acknowledgments}

This work is supported by NSFC under grant No.12505149.

\clearpage

\bibliographystyle{apsrev}
\bibliography{ref}{}

\end{document}